\documentclass[12pt,A4]{article}
\usepackage{epsfig}
\usepackage{array}
\usepackage{amsmath}
\usepackage{amssymb}

\newcommand{\m}{\mathbf}

\newcommand{\be}{\begin{equation}}
\newcommand{\ee}{\end{equation}}
\newcommand{\bea}{\begin{eqnarray}}
\newcommand{\eea}{\end{eqnarray}}

\title{Collinear true ternary fission as the consequence of the collective nuclear model }
\author{ F. F. Karpeshin \\ Mendeleev All-Russian Research
Institute of Metrology \\ 190005 Saint-Petersburg, Russia }

\begin{document}

\maketitle

\abstract{ The concept of collinear spontaneous true ternary   fission of $^{252}$Cf is subject to critical analysis. The conclusion is that the collinear flight of the fragments turns out to be a natural and most probable mode. The collinearity arises in the model on the prescission stage as a result of the account of the principles of the collective Bohr?s model. It is partly destroyed   at the post-scission stage of spreading of the fragments due to their Coulomb interaction, with the allowance for the spin effects arising  at the moment of scission.   The final angular distribution of the fragments is calculated by means of the  trajectory simulations. The  calculated relative angle of the heavy and light fragments is kept 180$^\circ$ with an uncertainty within 0.4$^\circ$, which justifies search for a collinear tri-partition at the modern stage of experiment. }

\vspace*{1cm}

\newpage

\large

\section{Introduction}

The question of fission into three comparable fragments has a long, challenging and fascinating story.  As distinct from traditional ternary fission, where emission of two massive fragments is accompanied with a ternary light particle, like an $\alpha$ particle, sometimes it is called true ternary fission (TTF).  Strutinsky {\it et al.} were the first who proposed search for this process \cite{stru}. Attempts of creating a theory of TTF were undertaken by many theorists. Proceeding from typical initial conditions on the top of usual fission barrier,  within the framework of the liquid drop model, Nix \cite{nix} demonstrated appearance of a third very light fragment which arose between two other massive fragments in the case of very heavy fissioning systems with $A\gtrsim 300$. Family of shapes leading to fission into three massive fragments was deduced in Ref. \cite{sierk,carj}. To this end, Legendre-polynomial expansion up to tenth order and more was exploited. In some papers, though, the role of hexadecapole deformation was underlined.  The idea of generic mechanism of TTF was expressed in Refs. \cite{coimbra,3fYaf,3fJ}, in contrast to the consecutive one. It is suggested that TTF develops along a special dynamical path to which the nucleus enters at the very beginning of fission. The predetermining role belongs to the hexadecapole deformation, like the quadrupole deformation plays the leading role in habitual binary fission.

      The first experimental searches for the effect of TTF were undertaken  in fission of actinide nuclei by thermal neutrons \cite{muga} and $\alpha$ particles \cite{iyer}. The measurements were also conducted   in heavy-ion collisions \cite{pere}, and spontaneous fission of $^{252}$Cf \cite{teo}. However, only upper limits of the probability of the processes were established at the level of 10$^{-4}$ -- 10$^{-8}$. It is worthy of noting that there was a tacit contradiction between theory and  experimental search.
From the theoretical point of view, the linear form of  fissile nuclei is more favorable than a clover-leaf shape ({\it e. g.} \cite{coimbra} and refs. cited therein).  However, experimental efforts were
mainly aimed at detecting fragments at approximately similar angles, i.e.,
$\sim 120^{\circ}$. Based on general considerations, the experimenters likely
believed that the mutual electrostatic repulsion could align the spreading angles.

	Solyakin {\it et al.} proposed the collinear mode of tripartition  \cite{solya}, when  searching for TTF of $^{238}$U by 1-GeV protons. This concept was most successfully realized in JINR experiments on FOBOS and mini-FOBOS setups \cite{fobos1,fobosROT,kama}. All the results presented below are obtained
within the framework of the ``missing-mass'' approach.
In fact, only two fragments were detected in each decay
event at a relative angle of 180$^{\circ}$. The mass and velocity of the ``missed'' fragment
were calculated based on the mass and momentum
conservation.
	The second principal feature of the spectrometer was the presence of the blocking grid, which 
prevented cases when light and ternary fragments could strike the same detector. 
Use of the missing-mass method in combination with
the supporting mesh on one of the  detectors led to conclusion that the collinear mode  of TTF of $^{235}$U and spontaneous fission of  $^{252}$Cf may be at the level of up to
10$^{-4}$ -- 10$^{-5}$. At first sight, this mode is in  contrast with ordinary ternary fission \cite{dbt}, where $\alpha$ particles or protons are emitted approximately perpendicularly to the fission axis. It is no coincidence that in Ref. \cite{doubt} the authors doubted a possibility of a ``perfectly'' collinear flight in the case of three massive fragments. At the same time, it is known that there is a small fraction $\sim$10 percent of polar alpha particles, emitted along the fission axis (e.g. \cite{polar} and refs. cited therein). Furthermore, direct estimations of possible scenarios of the fragment spread made in Refs. \cite{tash,izv,3fyaf} show that the collinear trajectory only can be expected if all the three fragments are born in one line. Minor displacement of the middle fragment from the axis $\sim$0.5 fm, as well as the presence of a non-zero transverse component of its initial velocity \cite{tash} break the collinearity. 
Therefore, the question of angular distribution of the fragments is of primordial interest.  

	As we will see, the collinear spreading of the fragments  follows the principles of  the A. Bohr's collective model of nuclear motion.  Consequently, the description turns out to be completely different  in the case of  traditional ternary fission,  accompanied by emission of an alpha particle, and TTF. In the former case, the two nascent massive fragments form an axially-symmetric core. It is  in the field of this core that the $\alpha$ particles are formed and then spread. In the case of TTF,  all the three nascent fragments form an axially symmetric nucleus. This means that they  move coaxially until separation. 

	The next question, however, arises whether this pre-scission collinearity may survive in the course of post-scission propagation of the fragments. 
They make a complex three-body motion under the action of the mutual Coulomb repulsion, at the same time keeping the memory of the rotation of the fission axis and that impulse they received at the moment of rupture. As it is shown in the next section, devoted to a qualitative consideration, combination of these factors destroys a collinear picture, but  to a some extent.
In section \ref{init}, formulas used for calculation are derived. The results of calculation for representative fragments of TTF are reported in section \ref{results}. They  are discussed in the concluding section.

\section{Qualitative premises}
\label{premises}

       As is known, projection  $K$ of the total angular momentum of a nucleus on the nuclear axis, related to the intrinsic nuclear coordinate system,  is a good quantum number. In quantum mechanics, the body cannot rotate around the axis of symmetry, and thus the rotational momentum is perpendicular to the axis of symmetry, and its projection $K$ onto the axis of symmetry is zero. At the same time, a nonzero value of $K$ can also be observed in an axially symmetric system due to quasiparticle excitations unrelated to rotation. If the system is a bit non-axially symmetric, then basically it will also rotate around an axis perpendicular to the axis of symmetry, and it will only  twist around the axis of symmetry slightly. In quantum mechanics, this is reflected in the fact that the wave function of the fissile nucleus has the following form \cite{BM}:
 \be
 \Psi^I_M (\m r_i) = \sum_K a_K D^I_{MK}(\theta, \phi, \vartheta)\chi_K(\m r'_i)\,.     \label{D}
 \ee
$I$ in Eq. (\ref{D}) is the nuclear spin, $M$ --- its projection in the laboratory frame. Wigner's $D$ functions from the Euler angles $\theta, \phi, \vartheta$ define the orientation of the nucleus in space and determine the angular distribution of the fragments.  $\m r_i$ and $\m r'_i$ are the nuclear variables ({\it e. g.}, nucleon coordinates) in the laboratory system and intrinsic coordinate system, respectively.  

	The $z'$ axis thus coincides with the fission axis. And let us direct  the $x'$ axis in the plane of symmetry of the fissile nucleus in the case of $K>0$. This may be a triangular configuration formed by three fragments that are not on the same line. Such a configuration was considered in \cite{tash,izv}. In terms of the Euler angles, the rotation from the laboratory system $\m r$ to the intrinsic system $\m r'$ can be performed in three steps. First two of them, the rotations by $\theta$ about the $z$ axis and by $\phi$ about the new axis $x'$, respectively,  impose the $z$ and $z'$ axes with each other \cite{BM}. After which, it remains the third rotation by $\vartheta$ about the new $z'$ axis, in order to impose the $x$ axis at the $x'$ one. In the case of a axially-symmetric nucleus, this third rotation evidently might not be  needed, as all the azimuthal angles are equivalent. The only way to combine this picture with Eq. (\ref{D})  is to put $K$ = 0. And {\it vice versa}: in the case of axially-asymmetric --- triangular configuration of the fissile nucleus, rotation by the angle $\vartheta$ is essential. Correspondingly, this excludes values of $K$ = 0 in Eq. (\ref{D}), leaving only values $K>0$ as adoptable. 

	Actually, this consideration is founded on the same arguments as the A. Bohr's hypothesis \cite{BM,BH} about the predominance of a certain channel in photofission of $^{238}$U.  As $I\geq K$, non-zero $K$ values are only possible if fissile nucleus has a non-zero spin. This is not the case if spontaneous fission of  $^{252}$Cf is considered. Therefore, it is only the symmetric configuration ``three in line'' which survives fission.   
	
	Furthermore, even such a ``co-axial'' initial configuration is not enough yet for the final collinearity. It can be destroyed during spreading of the fragments, as a result of interplay of the accelerating Coulomb force between them and initial velocity conditions. Most essential is perpendicular to the fission axis component of the initial velocity, which arises at the moment of scission due to big spins and  large relative angular momentum of the fragments. In the case of binary fission of actinide nuclei, the mean value of  the fragment spin   is about 
7 -- 8 \cite{Ras,Bosch}.
In papers \cite{Ras,nixvib},  appearance of the spins in the fragments at scission is explained by excitation of the collective modes of wriggling and bending vibrations. In the first case, the fragments are formed with spins parallel to each other, and perpendicular to the fission axis. Arising total spin  is compensated by the orbital angular momentum $L$ of the relative motion of the fragments. The latter is thus in the opposite direction to the total spin and also perpendicular to the fission axis, as shown in Fig. 1. The observed value of the mean spin of the fragments can be explained in this way (e.g. \cite{wrig}).

       One concludes from the above consideration that in the case of TTF, appearance of wriggling vibrations in all three fragments can lead to the total spin of the  fragments,  and, respectively, to their total orbital momentum  as much as $L\sim$ 20. And the larger the $L$ value, the greater the final angle of divergence between the fragments. Baring this in mind, we varied possible $L$ values within $0  \le L \le 20$. In the case of bending vibrations, the  fragments are formed with spins antiparallel to one another. Therefore,  the relative orbital momentum, together with the related destruction of the collinearity,  is expected to be even smaller.
       
        Allowance for the initial transverse velocity  results in the final divergence of the spreading fragments. If the fragments could move completely freely after scission, then with reasonable initial conditions, all the fragments would stay collinear on a rotating fission axis. The necessary condition for the fragments to remain on the axis is that both the transverse and longitudinal velocity components remain proportional to the distance from c. m.  When the Coulomb force is switched on, it changes only the longitudinal component. This violates the proportionality: the middle (ternary) fragment is pushed to the c. m. by the both outer fragments. This reflects, specifically, in its small final kinetic energy \cite{tash,izv,3fyaf}. In turn, the middle fragment itself pushes both the side fragments out, which also works as to violate the proportionality. 
As a result, as soon as the middle fragment descends from the axis, this immediately triggers the transverse component in the Coulomb repulsion between the fragments, which enhances the further destruction of collinearity. 
       
       Note that the mechanism described has much in common with the ROT effect, which arises  in fission of nuclei with spins, different from zero, by polarized neutrons \cite{goen,dan}. The ROT effect is 
the triple angular correlation between neutron spin and the momenta of fragments and ternary
particles.  The ROT effect can be explained as  due to rotation of the fission axis before scission, which is transferred to the fragments at scission as the transverse initial velocity. Naturally, this mechanism would also give a contribution to destruction of collinearity in TTF, if the fissile nucleus had a spin.   There is, however, a big difference in  the mechanism of ROT effect and that discussed above. First, the mechanism, destroying the collinearity in TTF,   appears to be  much stronger, as it is related with much higher momenta $L\lesssim 20$. Second, it does not contribute to the ROT effect because of angular averaging: there would be no correlation of the $L$ direction with the spin of the fissile nucleus, even if the spin were different from zero.    Contrary, in the case of  TTF, where each event is detected independently of  the others,   relative momentum $L$ could easily manifest itself through  violation of the collinearity. 

       To sum it up, we will assume that the total relative momentum of the fragments may reach as high as $L \approx 20$ in the case of wriggling vibrations. 
This value is compensated by the total spin of the fragments. 
The latter may be smaller or even zero in the case of bending vibrations. This will reply to  smaller  angular momenta $L$. Respectively, still more collinear trajectories of the fragments will be expected. Let us turn to the numerical estimations. \\

\section{Calculation formulas}
\label{init}

\subsection{Equations of motion}

      Numerical simulation of trajectories of representative fragments is a classical method. Its applicability follows a known fact that the wavelength related with  the fragment translation  is small as compared to its size. Such calculations were found to work well in description of the spectra and angular distributions of $\alpha$ particles, emitted in ternary fission (e. g.\cite{pias}), specifically, of the ROT effect \cite{gus}.

	Let the fission axis coincide with the quantization axis $z$ at the moment of scission. In view of the axial symmetry of the problem, let $x$ be the transverse direction axis. After scission, further trajectories of the fragments are determined by   the repulsive Coulomb forces between them. 

Denote  the side fragments with indices 1 and 2, and the middle fragment as No. 3 (Fig. 1). Representative trajectories are simulated in the next section by solving the Newton equations of motion for each of the fragments: 
\be
\frac{d^2\m r_i}{dt^2} = \m F_i / M_i, \qquad i = 1, 2, 3\,, \label{N}
\ee
where $M_i$ is mass of the $i$-th fragment, and $F_i$ is the resulting force acting on it from the two other  fragments. For simplicity, the latter is calculated under natural assumption of spherical fragments. 
System of coupled differential equations of the second order (\ref{N}) has to be solved  numerically with  the proper initial conditions concerning the positions of the fragments and their velocities at scission.

	\subsection{Initial conditions for the positions of the fragments} 

I consider  the generic mechanism of the TTF, when the both scissions occur nearly simultaneously.   The choice of the initial conditions is illustrated  in Fig. 1.  
\begin{figure}[!hbt]
\centerline{ \epsfxsize=15cm\epsfbox{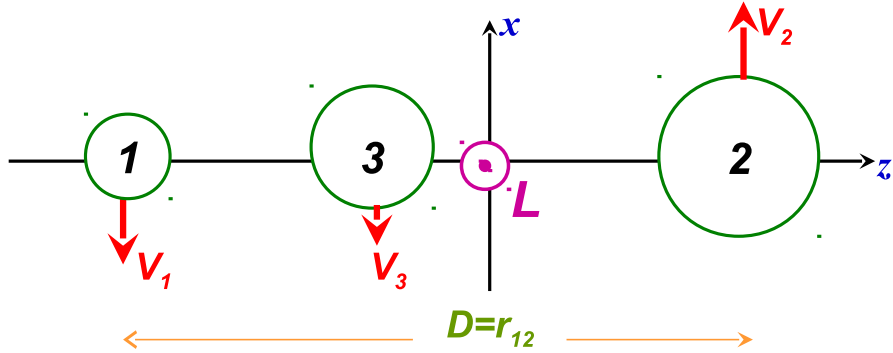}}
\vspace*{-4cm}
\caption{\footnotesize Initial conditions for the trajectory simulations. $V_1$, $V_2$ and $V_3$ are the transverse velocities to the fission axis of the fragments 1 -- 3, which comprise the total relative angular momentum of the collective rotation of the fragments (directed towards us). $D$ is the distance between extreme fragments. } \label{KinemF}
\end{figure}
And let us specify the atomic and mass numbers of the fragments as $Z_i$ and $A_i$, respectively,  with the distances $r_{12}$, $r_{23}$ and $r_{13}$  between the fragments. The positions of the fragments
must be defined, baring in mind their asymptotic total
kinetic energy (TKE), which must not exceed reaction
heat $Q$. For the parameterization purposes, the total
Coulomb energy of the fragments is minimized, based
on the position of the ternary fragment at fixed distance $D=r_{12}$ between the side fragments:
\be
r_{23}=D\frac{\sqrt{Z_2}}{\sqrt{Z_1}+\sqrt{Z_2}} \,.    \label{inz}
\ee
In this way, the initial positions of all three fragments are
fixed by the single parameter $D$, which in turn is  defined by the TKE value
$T$, $T\leq Q$:
\be
T = (\frac{Z_1Z_2}{r_{12}}+\frac{Z_1Z_3}{r_{13}}+\frac{Z_2Z_3}{r_{23}})\ e^2\, .
\label{iR}
\ee

\subsection{Initial conditions for the transverse velocities of the fragments} 
\label{inveloc}

     In the laboratory system, the most general motion of  the fragments can be represented as a superposition of  a  linear translation and rotation  around their center of  mass.  The former gives linear velocity of the fragment, the latter is nothing more than the spin of the fragment. 
       A small initial velocity of the fragments in the  direction of fission is not important for the present purposes. In order to calculate the initial transverse velocity of the fragments, let us designate the masses of the fragments and their positions on the axis of fission as $M_1$, $z_1$, $M_2$, $z_2$  and $M_3$, $z_3$, respectively. The center of gravity of the fragments, determined during fission, is set as
\be
\zeta=(M_1z_1+M_2z_2+M_3z_3)/M \,,
\ee
where the total mass $M=M_1+M_2+M_3$.
The total angular momentum of the fragments $L$ is defined as follows:
\be
\omega\ [M_1(z_1-\zeta)^2+ M_2(z_2-\zeta)^2+ M_3(z_3-\zeta)^2]=L\hbar  \, ,
\ee
and the initial transverse velocity of fragment $i$  is
\be
V_i=\omega (z_i-\zeta)\,.   \label{invel}
\ee

\section{Results of the calculation }
\label{results}

A landscape of the potential deformation energy was calculated in Ref. \cite{karpov} for the case of TTF of $^{252}$Cf. It suggests the following  mode as a likely candidate:
\be
\text{$^{252}$Cf} \to  \text{$^{132}$Sn}+\text{$^{48}$Ca}+\text{$^{72}$Ni},
\qquad Q = 251 \, \text{MeV}  \,,  \label{c1}
\ee
with the light and heavy side fragments of Ni and Sn, and the ternary fragment of Ca in the middle.
The $Q$ value in fission (\ref{c1}) was calculated, using  AME2012 atomic mass evaluation \cite{AME}. The presence of two magic or semi magic fragments in the final state provide a great released energy $Q$. The situation is like in three-partition of the atomic clusters  of  $_{27}Na^{+++} \to 3\ _9Na^+$ into three magic clusters of $_9Na^+$  \cite{coimbra}. The final TKE values depend on the  scission configuration: position of the fragments, thickness of the necks. Deformation of the fragments takes a part of energy  from the $Q$ value, diminishing TKE of the fragments.  For this reason, I consider various representative TKE values and total angular momenta $L$.  

       In the  landscape of the potential energy  \cite{karpov}, pronounced valleys favorable for ternary fission were found.  One of them, which may be related with channel (\ref{c1}), lies after a saddle point at $r_{12}\approx 3R_0=$ 22 fm,    where $R_0$ is the radius of the mother nucleus.
At this distance, formation of the future fragments starts.  The valley presents a good opportunity for scission and separation of all three fragments somewhere at $r_{12}\gtrsim$ 30 fm. Indeed, the TKE value $T=Q$ would be achieved if scission occurred at $r_{12}$ = 25.56 fm. In practice, part of the released energy is stored in the deformation energy of the fragments, while scission occurs at a larger distance.  Baring this in mind, we varied the parameter $D=r_{12}$ in the range up to $D$ = 40 fm. Experimental results \cite{kama} confirm such an  expectation.
      
       Equations of motion (\ref{N}) with initial conditions (\ref{inz}) and (\ref{invel}) were solved by means of the Runge---Kutta---Nystr{\"o}m method. 
The results of the trajectory simulations  are presented in Fig. 2 and Table 1.  The calculated kinetic energies of each of the fragments, together with their TKE, are presented in Fig. 2 {\it versus}  the distance between the side fragments $D$ at scission. 
	All the energies smoothly  decrease with increasing $D$, while the TKE changes from $T$ = $Q$ = 251 MeV for $D$ = 25.6 fm down to $T$ = 160 MeV for $D$ = 40 fm. As well as in ordinary binary fission, the heavy fragments are produced with lower kinetic energies.  We note a characteristic feature of TTF, which follows Fig. 2:  the ternary fragments, which are formed between the heavy and light ones, turn out to be very  slow, with the kinetic energies of approximately 5 MeV. This is 15 -- 20 times as small as the energies of the main fragments.  Such low energies are in accordance with refs. \cite{tash,3fyaf,erzen}. Qualitative reason is that in a collinear flight, the motion of the ternary fragment is confined by the two outer fragments.  \begin{figure}[!bt]
\centerline{ \epsfxsize=15cm\epsfbox{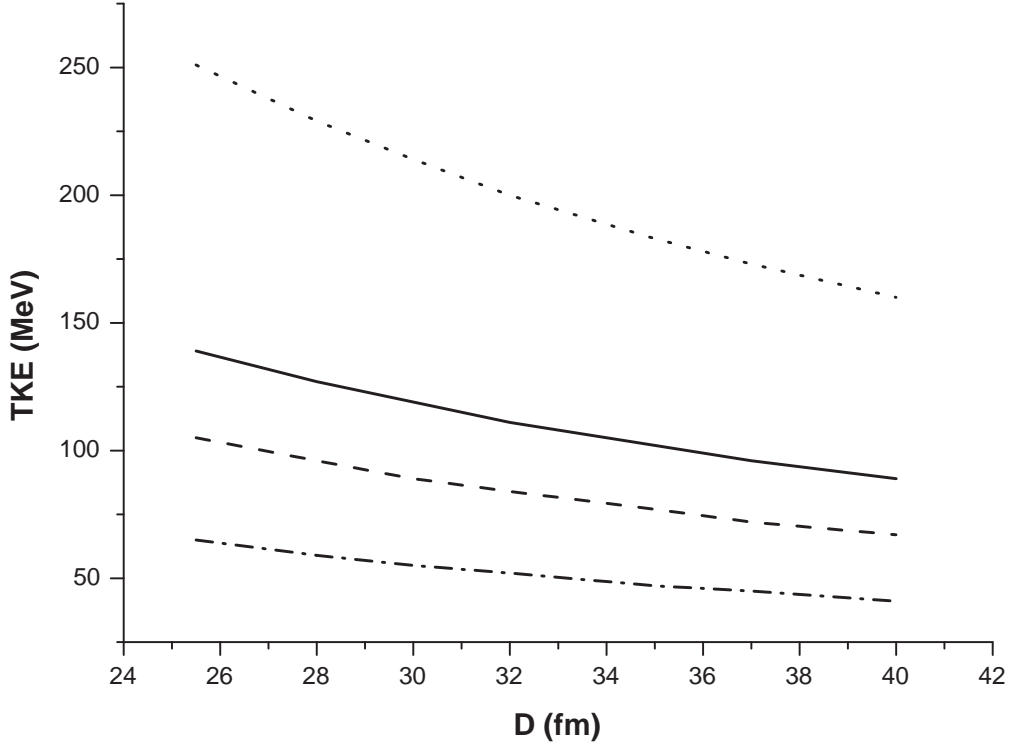}}
\caption{\footnotesize Kinetic energies of the fragments and their TKE values against the scission point $D$: main fragments of $^{72}$Ni and
$^{132}$Sn --- full and dashed lines, respectively, dash-dotted line --- the energy of the ternary fragment of $^{48}$Ca, scaled by a factor of 10,  and dotted line --- the total kinetic energy.} \label{KinEnF}
\end{figure}

      Results concerning the angular distribution of the fragments are presented in Table 1.
As a consequence of the rotation of the fission axis, none of the fragment trajectories remains on the $z$ axis after scission, if $L \neq 0$. For the configuration presented in Fig. 1, where the momentum $\m L$ is aimed at the reader, the light Ni fragment goes down from the $z$ axis. In turn, the heavy Te fragment comes up.
As a result, the two main  fragments scatter in the opposite directions, with the angle $\Theta$ between them remaining close to 180$^{\circ}$. The calculated $\Theta$ values are  displayed in the Table for various momenta $L$. As one can see from the Table, the difference from 180$^{\circ}$ does not exceed 0.4$^{\circ}$. This deflection is a consequence of the transverse force, discussed in Section \ref{premises}. 
The calculated  angles between the light and heavy fragments satisfy the experimental conditions \cite{fobos1,fobosROT,kama}, where only collinear fission events with a relative angle of $180^\circ\pm 2^\circ$ were selected.

	The ternary fragment always flies in the same direction as the light one. The angle of divergence $\Phi$ between them is also presented in the Table against   the total  angular momentum $L$. The latter was  varied in a wide range $0\le L \le 20$, as explained previously.  All three fragments remain in the same plane. Projections of the velocities of the light and ternary fragments  on the axis, perpendicular to the direction of the heavy fragment, have opposite signs. Scattering of the ternary fragment into the upper and lower half-planes in Fig. 1  is equally probable. The results presented clearly show that the Ni and Ca fragments, moving in the same direction, diverge within
one-two degrees at most, for all the considered $L$ values. However, with the energies, presented in Fig. 2, the ternary fragment arrives at the range of the FOBOS or mini FOBOS  detector with delay of $\sim 10^{-7}$ s as compared to the light fragment. 
\begin{table}[bth]
\caption{\footnotesize Calculated angular distributions of the fragments of true ternary fission of $^{252}$Cf (\ref{c1}) {\it versus} the scission distance $D$ and the relative angular momentum $L$.  $\Theta$ is the angle between the directions of the heavy and light fragments,
$\Phi$ --- the divergence angle between the light $^{72}$Ni and ternary  $^{48}$Ca fragments, moving in the same direction}
\begin{center}
\begin{tabular}{||c| c||c|c||}
\hline    \hline
$D$, fm   &   $L$   & $\Theta^{\circ}$  & $\Phi^{\circ}$    \\
\hline
&  5  &  179.9  &  0.7   \\
\raisebox{-15pt}[0pt][0pt]{25.6} &    10  &  179.8  &  1.4  \\
   &  15  & 179.7   &  2.1   \\
  &  20  & 179.6  &     2.8 \\
\hline
\raisebox{-15pt}[0pt][0pt]{30} &      5  &  179.9  &   0.6  \\
  &  15  &  179.7  &  1.9   \\
\hline
  &  5  &  179.9  &  0.6   \\
35  &  15  &  179.7  &   1.8     \\
   &  20    &  179.6  &  2.4     \\
\hline  \hline
\end{tabular}
\end{center}
\end{table}

\section{Discussion}

It follows from the considered model that the light and heavy fragments fly in the opposite directions with the relative angle 180$^\circ$ with the accuracy of 0.1$^\circ$ -- 0.4$^\circ$. This is in contrast with the first experiments \cite{muga} and others, aimed at detecting fragments at the angles of $\sim 120^\circ$ to one another. The above result thus justifies search for 
collinear mode  of TTF  as most probable. The result obtained bases on the two circumstances. The first circumstance   is  axial symmetry  of the fissile nucleus on its path towards scission. Because of the axial symmetry, there is a sole way of formation of the fragments, when they remain co-axial till scission. This is a remarkable illustration of the collective A. Bohr's model. Furthermore, it is in spirit of the A. Bohr's hypothesis about predominance of the channel with $K$ = 0 in photofission of $^{238}$U. This hypothesis transfers the principles of symmetry from the collective model into fission. As is known, the A. Bohr's hypothesis also works in the case of fission of $^{235}$U by thermal neutrons, in which case the compound nucleus is characterized by 
strong mixing of the states with different   $K$ values, so as all the possible  values from 0 to 4 may become nearly equally probable due to the Coriolis mixing \cite{fur}. As a result, it was suggested that the channel  $J^\pi K =4^-0$ should also be considered for the resonances with  $J^\pi=4^-$. Also in this case, most probable channels with a certain  $K$ reply to minimal energy over the fission barrier. As an example of application of the symmetry principles in fission, we also note paper \cite{fur2} where, in addition to the $J^\pi K$ characteristics of fission barriers, the signature quantum numbers $s$ and $r$ related to the symmetry
of the first and second fission barriers were introduced, aimed at further studying the properties of fission of nuclei with $K$ = 0.   One can say that the barrier works as the filter, which selects the channels with minimum energy over the barrier.  However, the angular distribution in the case of fission of nuclei with spins becomes more complicated  because of the strong mixing  over $K$ values on the stage of the compound nucleus \cite{fur}. A consideration of  such cases can be performed elsewhere, in a similar approach. 

      The second circumstance is good conservation of the collinearity during the post-scission spreading of the fragments under the action of the Coulomb repulsive force. The repulsion leads to descent of the fragments from the axis, after which appearance of the transverse component of the Coulomb force forms the final value of the angles of divergence between the fragments. The  final angle  between the light and ternary fragments  remains at the level of two degrees. The kinetic energy of the ternary particle turns out to be very low, around 4 -- 7 MeV.

		In the trajectory simulations, I did not take into account the strong interaction  of the fragments, whereas in \cite{tash} the authors did.
At first glance, this may seem to be a shortcoming. However, at deeper insight, my present model is more realistic, than used in \cite{tash}. Paper \cite{tash} contains many physical errors, among which I note the following.

1) Unrealistic initial configuration of three touching spheres, without attention to the dynamics of fission.  The distance between the side fragments $R_{12}$  = 20 fm is rather typical for binary fission of  a smaller actinide nucleus, such as $^{238}$U. In that case, one has to take into account that before scission, a long neck is formed. After its rupture, the remnants of the neck  snap back \cite{halp}.       Thus,    in the  landscape of the potential energy  \cite{karpov}, pronounced valleys favorable for ternary fission were found.  One of them, which may be related with channel (\ref{c1}), lies after a saddle point at $r_{12}=$ 22 fm.  The TKE value $T=Q$ would be achieved if scission occurred only at $r_{12}$ = 25.56 fm. In practice, part of the released energy is stored in the deformation energy of the fragments, while scission occurs at a larger distance.

2)  And only after this, the fragments can be considered as approximately spherical, being at a considerable  distance from one another, where short-range strong interaction really does not play a noticeable role. Choice of the initial configuration in the present work: $R_{12}$  = 27 -- 35  fm is in agreement with the above picture and microscopic dynamical calculation \cite{karpov}. 

3) Moreover, one expects that the configuration \cite{tash} will rather lead to the merging of the three spheres into one mother nucleus, than to their further separation. 
 
4) Not to mention, that there are tens versions of the effective strong forces in the nuclei in the literature, and making use of each of them will yield in different results, including absurd. In the case of any realistic scenario of fission, this does not make any problem. But it bares problems in the case of extreme scenario, like in \cite{tash}. For example, if more advanced potential from \cite{NPpyat} is used, the total energy of the configuration will exceed the Q-value by 15 MeV. 

5) Even with the interaction described in Ref. \cite{tash}, the total energy of the considered configuration exhausts the total  Q value for the given channel, so that the total kinetic energy (known as TKE)  of the fragments  will correspond to the case of so-called cold fission, never seen in double or triple fission.

6) Another absurdity arises when considering the initial velocity in \cite{tash}. The authors in \cite{tash} do not write, how to combine availability of the transverse initial velocity $v_{3y}$ on the middle fragment with the momentum conservation, if the other velocities are zero:  $v_{1x}=v_{2x}=v_{3x}=v_{1y}=v_{2y} = 0$.  My  model, presented in Section 3.3, harmonically satisfies the momentum conservation.

7)  The attention is drawn to a long meaningless chain of mathematical formulas, the result of which is... the derivation of the Newton's equations, starting from (3), (4) to (6), (7) in \cite{tash}, in the framework of the Lagrangian formalism. I am forced to remind you that the Newton's laws, being such, do not need to be justified.

   The first error seems to be the heaviest and fatal one. Even mistake No. 6, though related with violation of a fundamental law, can be understood as a consequence of incomplete thinking. But so much was written e.g. about non-reversibility of fission with fusion channel, which arises because of absence of the formations like remnants of the neck in the fusion channel. The authors write about dynamics in their calculation. But the dynamics in fission occurs at the prescission stage. Cf. e. g. classical works by Swiatecki, Nix, Koonin, Sierk {\it et al.} \cite{nix,sierk,carj} and others. Therein, calculation of the dynamics starts on the saddle, not at scission. The authors \cite{tash} also forgot about scission dynamics, model of which is proposed herein for the first time. Regretably, error No. 1 makes evidence of the generation gap. 

   Dwelling on these and other shortcomings of \cite{tash} is not among my present purposes, although. The main difference is that the principal question: collinear or not collinear case will be realized, actually remains unanswered in \cite{tash}.  I want to show the deeper physical reason, which makes collinear TTF most probable. 
It is remarkable that the final collinearity of the fragments gives a strong evidence  of the ideal axial symmetry  of the fissile system in its evolution up to  scission. This collinearity  would be already  broken by the very minimum shift of the middle fragment from the fission axis by $\lesssim$0.1 fm \cite{izv}. This comprises as little as 0.1 percent of the distance between the side fragments. This makes TTF a unique process showing manifestation  of the axial and other symmetries which underlie the A.  Bohr's model. Although  the collective model was designed for description of other phenomena, such as spectra and intensities of gamma quanta, {\emph etc.},  in none of these examples does the description  achieve  such an accuracy, may be, 
$\sim$10 percent at most. Therefore, TTF turns out to be a process where the merits and the underlying symmetries of  the collective model manifest themselves in full shine.

\section{Acknowledgments}

The author is grateful to L. F. Vitushkin, Yu. I. Gusev, and I. S. Guseva for their helpful advice, as well as to V. E. Bunakov, V. I. Furman, S. G. Kadmensky and A. V. Karpov for fruitful comments. He
also thanks D. V. Kamanin and Yu. V. Pyatkov for discussions of the experimental data.


\end{document}